\documentclass{PoS}

\usepackage{bm}% bold math
\usepackage{mathptmx}
\usepackage{amsmath}
\usepackage{bm}
\usepackage{textcomp}
\usepackage{todonotes}

\DeclareMathAlphabet\mathbfcal{OMS}{cmsy}{b}{n}

\title{Improved measurements of the energy and shower maximum of cosmic rays with Tunka-Rex}

\ShortTitle{Tunka-Rex: Improved measurements of the energy and shower maximum}

\author{
\speaker{D.~Kostunin}$^{1}$, P.A.~Bezyazeekov$^{2}$, N.M.~Budnev$^{2}$, D.~Chernykh$^{2}$, O.~Fedorov$^{2}$ O.A.~Gress$^{2}$, A.~Haungs$^{1}$, R.~Hiller$^{1}$\thanks{now at the University of Z\"urich}, T.~Huege$^{1}$, Y.~Kazarina$^{2}$, M.~Kleifges$^{3}$, E.E.~Korosteleva$^{4}$, O.~Kr\"omer$^{3}$, L.A.~Kuzmichev$^{4}$, V.~Lenok$^{1}$, N.~Lubsandorzhiev$^{4}$, T.~Marshalkina$^{2}$, R.R.~Mirgazov$^{2}$, R.~Monkhoev$^{2}$, E.~Osipova$^{4}$, A.~Pakhorukov$^{2}$, L.~Pankov$^{2}$, V.V.~Prosin$^{4}$, F.G.~Schr\"oder$^{1}$, A.~Zagorodnikov$^{2}$
--
Tunka-Rex~Collaboration \\
\llap{$^1$} Institut f\"ur Kernphysik, Karlsruhe Institute of Technology (KIT), Karlsruhe, Germany\\
\llap{$^2$} Applied Physics Institute of Irkutsk State University, Irkutsk, Russia\\
\llap{$^3$} Institut f\"ur Prozessdatenverarbeitung und Elektronik, Karlsruhe Institute of Technology (KIT), Karlsruhe, Germany\\
\llap{$^4$} Skobeltsyn Institute of Nuclear Physics MSU, Moscow, Russia\\
E-mail: \email{dmitriy.kostunin@kit.edu}       
}

\abstract{The Tunka Radio Extension (Tunka-Rex) is an array of 63 antennas located in the Tunka Valley, Siberia.
It detects radio pulses in the 30-80 MHz band produced during the air-shower development.
As shown by Tunka-Rex, a sparse radio array with about 200 m spacing is able to reconstruct the energy and the depth of the shower maximum with satisfactory precision using simple methods based on parameters of the lateral distribution of amplitudes.
The LOFAR experiment has shown that a sophisticated treatment of all individually measured amplitudes of a dense antenna array can make the precision comparable with the resolution of existing optical techniques.
We develop these ideas further and present a method based on the treatment of time series of measured signals, i.e. each antenna station provides several points (trace) instead of a single one (amplitude or power).
We use the measured shower axis and energy as input for CoREAS simulations: for each measured event we simulate a set of air-showers with proton, helium, nitrogen and iron as primary particle (each primary is simulated about ten times to cover fluctuations in the shower maximum due to the first interaction).
Simulated radio pulses are processed with the Tunka-Rex detector response and convoluted with the measured signals.
A likelihood fit determines how well the simulated event fits to the measured one.
The positions of the shower maxima are defined from the distribution of chi-square values of these fits.
When using this improved method instead of the standard one, firstly, the shower maximum of more events can be reconstructed, secondly, the resolution is increased.
The performance of the method is demonstrated on the data acquired by the Tunka-Rex detector in 2012-2014.
}

\FullConference{35th International Cosmic Ray Conference -- ICRC2017\\
		10-20 July, 2017\\
		Bexco, Busan, Korea}

\begin{document}

\section{Introduction}
Recent studies~\cite{Schroder:2016hrv} have shown that we understand the radio emission from air-showers created by ultra-high energy ($> 100$~PeV) cosmic rays on a level of better than 20\%.
Theoretical calculations show, that the resolution of a radio detector operating in frequency band of 30-80~MHz under ideal conditions can be 4\% for the electromagnetic energy~\cite{Glaser:2016tng} and 11~g/cm\textsuperscript{2} for the depth of shower maximum~\cite{Zilles:2017emn},
Meanwhile experiments give a resolution of 15\%~\cite{Bezyazeekov:2015ica} and 20~g/cm\textsuperscript{2}~\cite{Buitink:2014eqa}, respectively, and one has to keep in mind, that the last result is obtained with small, but dense array with statistical sensitivity limit to few hundreds PeV.

Tunka-Rex, a radio detector deployed in 2012 (for details see Ref.~\cite{Bezyazeekov:2015rpa,Schroeder_TunkaRex_ICRC2017}), is a sparse array suited for the detection of ultra-high energy cosmic rays using the frequency band of 30-80~MHz.
Up to now, for the reconstruction of air-shower parameters (primary energy and depth of shower maximum) Tunka-Rex used a method based on the LDF (lateral distribution function)~\cite{Kostunin:2015taa}.
By this method we performed a first direct cross-check between reconstructions provided by radio and optical (air-Cherenkov) measurements demonstrating a resolution of Tunka-Rex of 15\% for the primary energy and 40~g/cm\textsuperscript{2} for the depth of shower maximum.
The idea of this method was based on the basic features of the radio emission: geomagnetic and Askaryan emissions, and an exponential LDF (see Refs.~\cite{Kostunin:2016xbk,Kostunin:2015taa} for details).
Particularly, the energy is reconstructed as proportional to the amplitude at 120~m axis distance and the distance to the shower maximum is proportional to the slope of LDF at 180~m axis distance.
%~ Although it shows good performance, many things can be improved further.
%~ Particularly, the signal processing was performed in a very simple way, make possible the small fraction of false positive of misreconstructed signals, 
%~ then only amplitudes were taking into account (neglecting pulse shape).
%~ Finally, the lateral distribution was fitted with few parameters neglecting higher-order features.

In the present work we develop a completely new analysis method for the Tunka-Rex experiment, which combines sophisticated signal processing using matched filtering and the statistical top-down approach developed by LOFAR~\cite{Buitink:2014eqa}.
Similar to LOFAR, our method requires a set of CoREAS~\cite{Huege:2013vt} simulations to cover the range of possible shower-to-shower fluctuations.

\section{Initial dataset and simulation procedure}
Since 2015, Tunka-Rex has two different types of trigger:
by dense air-Cherenkov array Tunka-133 operating during winter moonless nights,
and Tunka-Grande scintillator array operating the rest of the time.
As first step, only events triggered by the air-Cherenkov detector Tunka-133 in 2012-2014 are selected for this work
which brings two advantages compared to Tunka-Grande triggered events.
First, Tunka-133 features 6~m resolution for the core reconstruction~\cite{Prosin:2016jev}, while the combined resolution of Tunka-Grande and Tunka-Rex is only about 20~m~\cite{Kostunin:2017bzd} since Tunka-133 is denser.
Having worse resolution, the core would have to be included as free parameter, which would make computations (simulations) more complex and introduce additional systematics.
Second, Tunka-133 reconstructs the depth of shower maximum and primary energy with precisions of about 28~g/cm\textsuperscript{2} and 10\%, respectively~\cite{Prosin:2016jev} which can be used to estimate the resolution of the updated Tunka-Rex analysis by cross-comparison.
Moreover, Tunka-133 reconstruction for 2015-2016 is kept in secret for the further cross-calibration with Tunka-Rex.

To obtain the initial configuration set, the events of the standard Tunka-Rex analysis~\cite{Bezyazeekov:2015ica} are taken, i.e. the simulations are configured with
the shower core reconstructed by Tunka-133 and the primary energy reconstructed by Tunka-Rex.
CoREAS simulations are produced using different primary particles: protons, helium, nitrogen and iron nuclei.
The exact Tunka-Rex layout relevant for each measured event is used, i.e. for the events measured in 2012-2014 the signals at a maximum of 25 antenna stations are simulated.
The number of simulations per event is selected dynamically in order to cover the possible shower-to-shower fluctuations, in other words, to cover the range of possible values of shower maxima for the energy of the event.
Thus, about 50 simulations per event are produced.

\section{Signal processing}
CoREAS, a simulation software for radio emission from air-showers, calculates the electrical-field vector at each Tunka-Rex antenna station.
These electrical fields are convoluted with the inverted hardware response to obtain the simulated ADC counts in every channel using the Offline software~\cite{Abreu:2011fb}.
After this, the measured and simulated ADC counts are convoluted with the direct hardware response of Tunka-Rex and digital filters from the standard analysis.
The resulting signal traces are upsampled by a high factor (16 in our case) to obtain sub-ns sampling (${16\times 200}$~MS/s) for precise determination of the peak time.

The coordinates of reconstructed electrical fields are converted to the geomagnetic coordinate system (see Ref.~\cite{Kostunin:2016xbk}), and only the strongest component, namely the ${\mathbf{V}\times\mathbf{B}}$ component, is considered in this proceeding.
Let us define the measured and simulated traces as $E_\mathrm{m}$ and $E_\mathrm{s}$, respectively.

\begin{figure}
\centering
\includegraphics[width=0.49\linewidth]{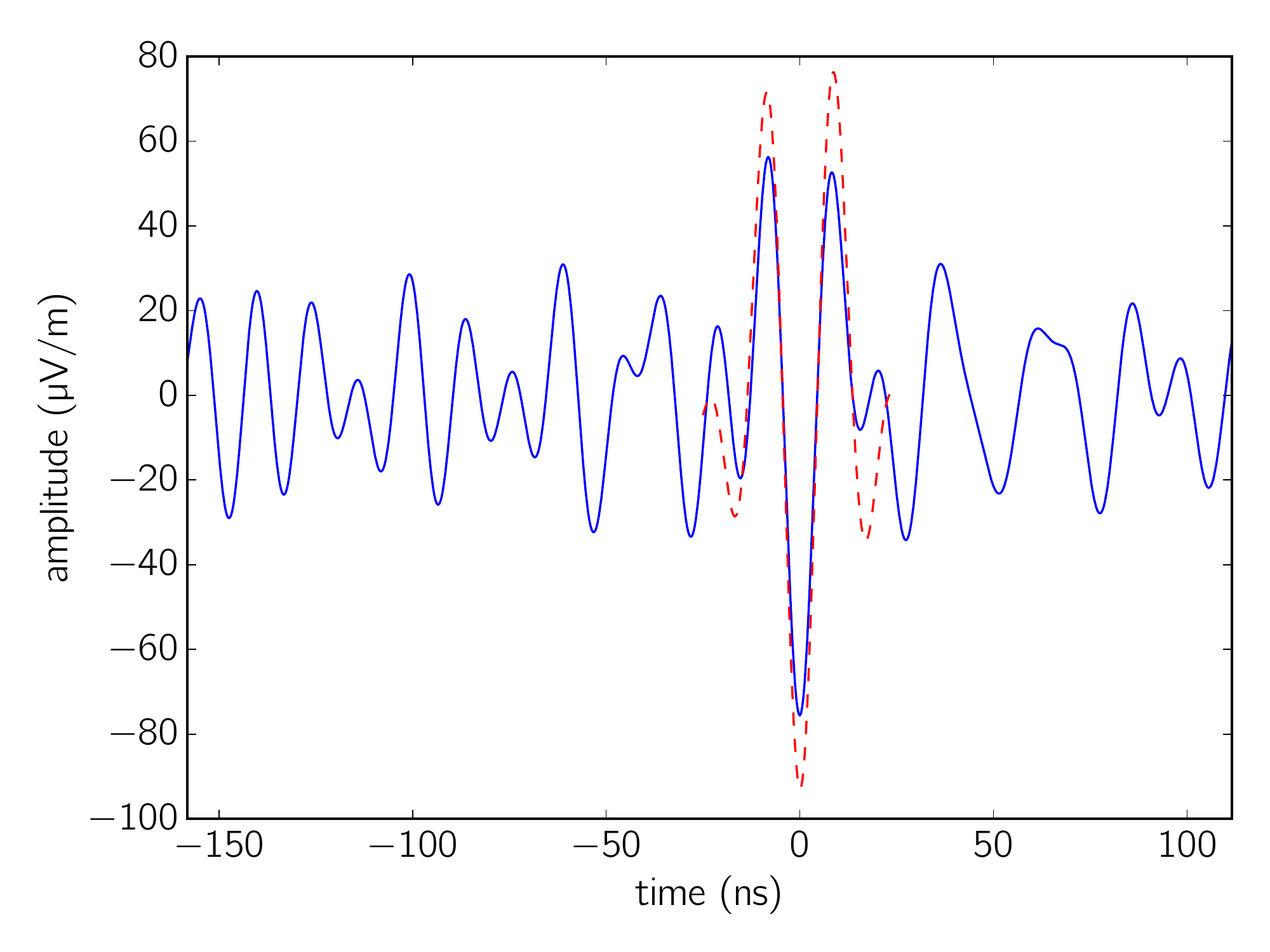}~\includegraphics[width=0.49\linewidth]{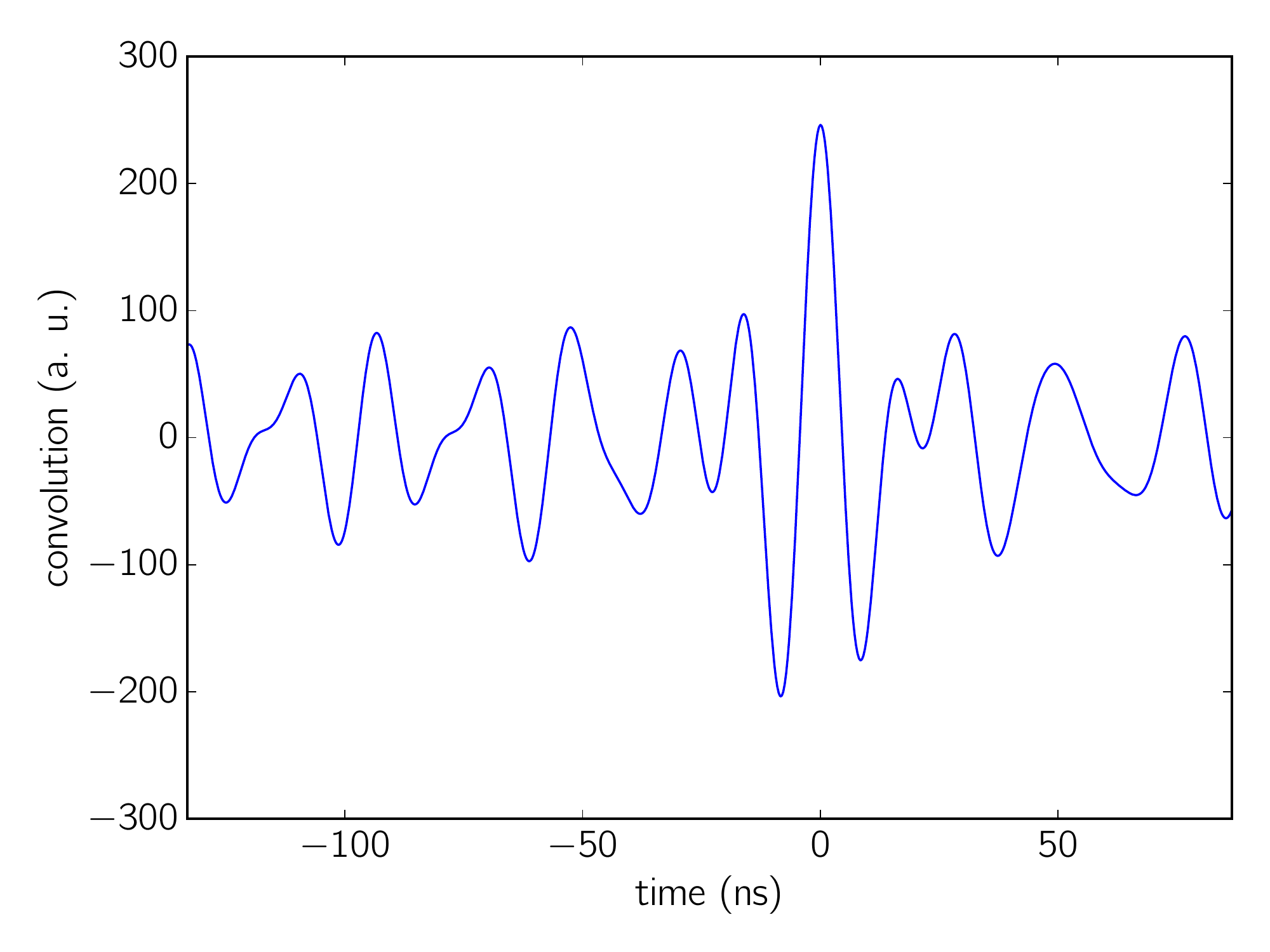}\\
\caption{
\textit{Left:} Tunka-Rex trace $E_\mathrm{m,signal}$ (solid) in signal window in comparison with simulated signal $v(t)$ (dashed).
\textit{Right:} Convolution of measured and simulated signals.
The peak of the left and maximum of the right timeseries are shifted by $t_0$ to be centered around zero for illustration purposes.
One can note that the pulse shapes are in best agreement in $\pm 15$~ns around peak, which is taken as window for further analysis.
}
\label{fig:tc}
\end{figure}

By design, the simulated trace $E_\mathrm{s}$ contains no background and the peak inside it can be easily found
Then the subtrace with $\pm 25$~ns around peak is taken as signal template $v(t)$.
Finding a peak in a measured trace is a more complicated problem.
Due to systematic uncertainties in timing and distortion by background, the position of the measured signal is unknown, moreover the full trace can contain RFI peaks, which might be recognized as false positives.
To define the signal peak and its signal-to-noise ratio (SNR) there are two windows selected in $E_\mathrm{m}$ similar to the standard analysis: a signal window $E_\mathrm{m, signal}$ and a noise window $E_\mathrm{m, noise}$.
Thus, we apply matched filtering to $E_\mathrm{m, signal}$ and define the peak time as $t_0$.
Fig.~\ref{fig:tc} shows an example of such filtering.
For the further processing (fitting) we decided to consider the filtered signals $u(t)$ with a width of $t_w = 30$~ns around the pulse peak, since broader timeseries are much affected by background, i.e. after matched filtering the template signal $v(t)$ is also shortened to this window.

After the filtering and peak determination the following quality cuts are applied for every filtered signal $u(t)$:
\begin{itemize}
\item \emph{SNR cut}.
	We define the signal-to-noise ratio as 
	\begin{equation}
	\mathrm{SNR} = \left(\frac{\max(E_\mathrm{m, signal} \ast v)}{\mathrm{RMS}(E_\mathrm{m, noise} \ast v)}\right)^2\,, 
	\end{equation}
	and select measured signals, which have ${\mathrm{SNR} > 6}$ for every simulated signal.
\item \emph{Peak drifting cut}. 
	This is a very strong and useful cut preventing false positives.
	Let us have peak times ${t_1,\,\,t_2,\,\,...,\,\,t_N}$ given after matched filtering of $N$ simulated signal templates.
	For this set we apply the following cut: $\forall i,j \le N : t_i - t_j < 5$~ns.
	In other words, each simulation must predict the same peak in the trace, otherwise the peak considered as false positive and the antenna station is rejected for the complete analysis of the event.
\end{itemize}

\section{Reconstruction of air-shower parameters}
\begin{figure}[t]
\includegraphics[width=0.49\linewidth]{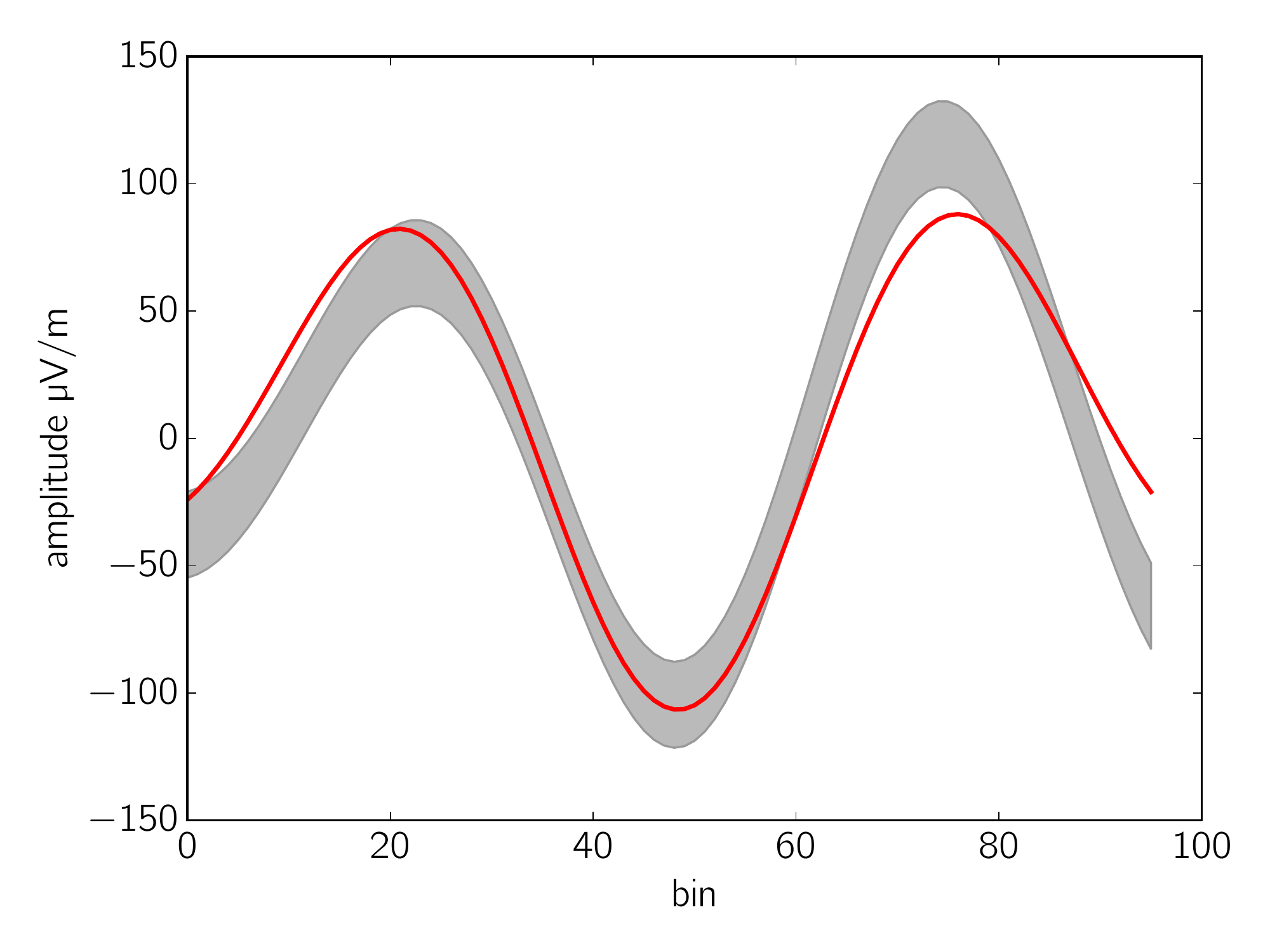}~\includegraphics[width=0.49\linewidth]{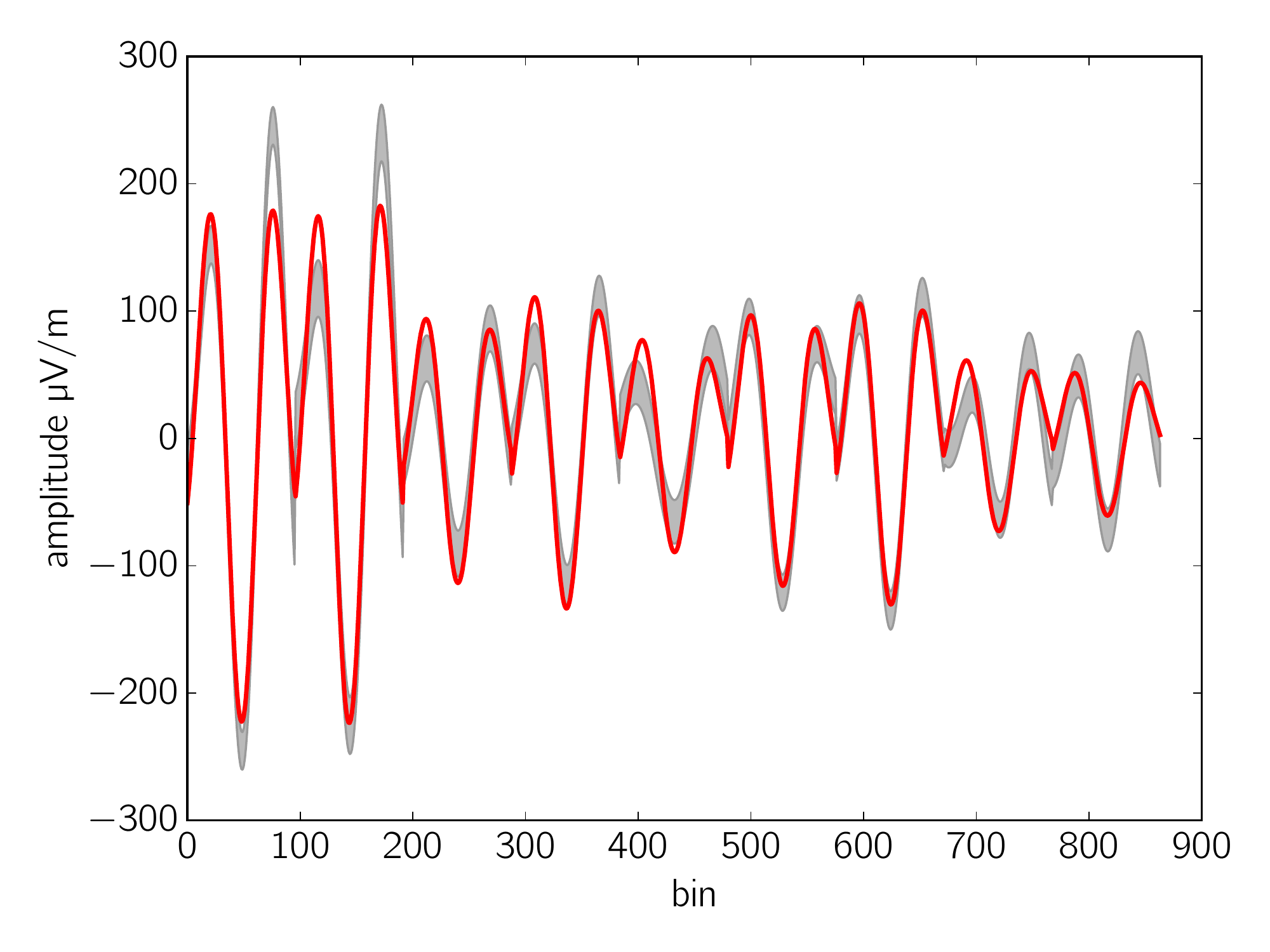}\\
\centering
\caption{Example of the measured traces $u(t)$ (shaded area includes uncertainties $\sigma$) and the best fit $Av(t)$.
\emph{Left:}~Example of event containing only one station with signal.
\emph{Right:}~Example of an event with the maximum number of stations (nine).
Traces are concatenated for simplicity.
}
\label{fig:traces}
\end{figure}

Similar to the LOFAR approach~\cite{Buitink:2014eqa} for each simulation we minimize the following $\chi^2$:
\begin{equation}
\chi^2 = \sum\limits_{i = 1}^N\sum\limits_{t}\left(\frac{u_i(t) - Av_i(t)}{\sigma}\right)^2\,,
\end{equation}
where $i$ is the index over stations with signals, the measured $u(t)$ and simulated $v(t)$.
$t$ is the bin in the signal timeseries, 
$A$ is free parameter for normalization,
$\sigma$ is the RMS of noise (defined in noise window $E_\mathrm{m, noise}$).
Two examples of the fit is given in Fig.~\ref{fig:traces}.

\begin{figure}[t]
\centering
\includegraphics[width=0.49\linewidth]{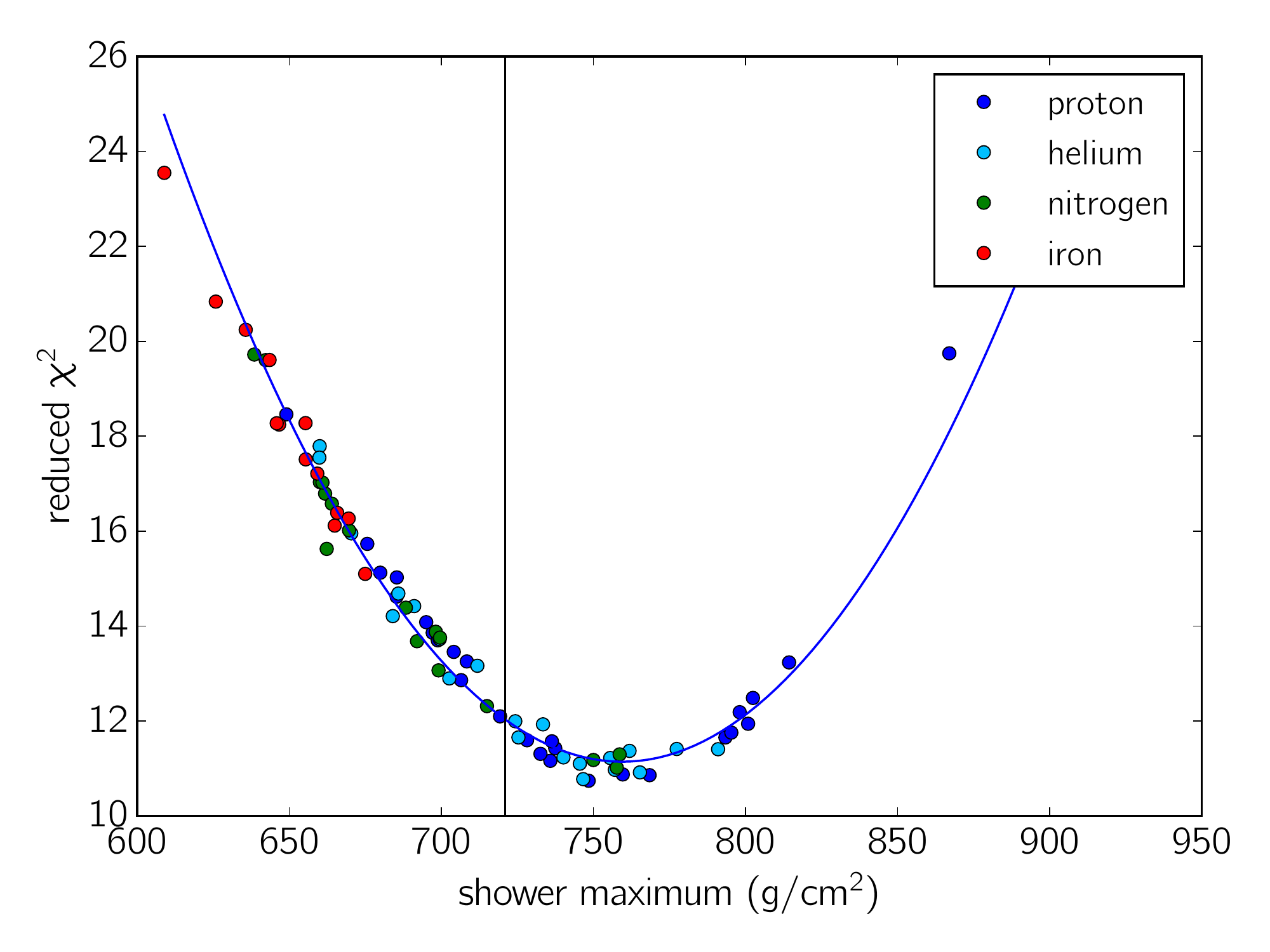}~\includegraphics[width=0.49\linewidth]{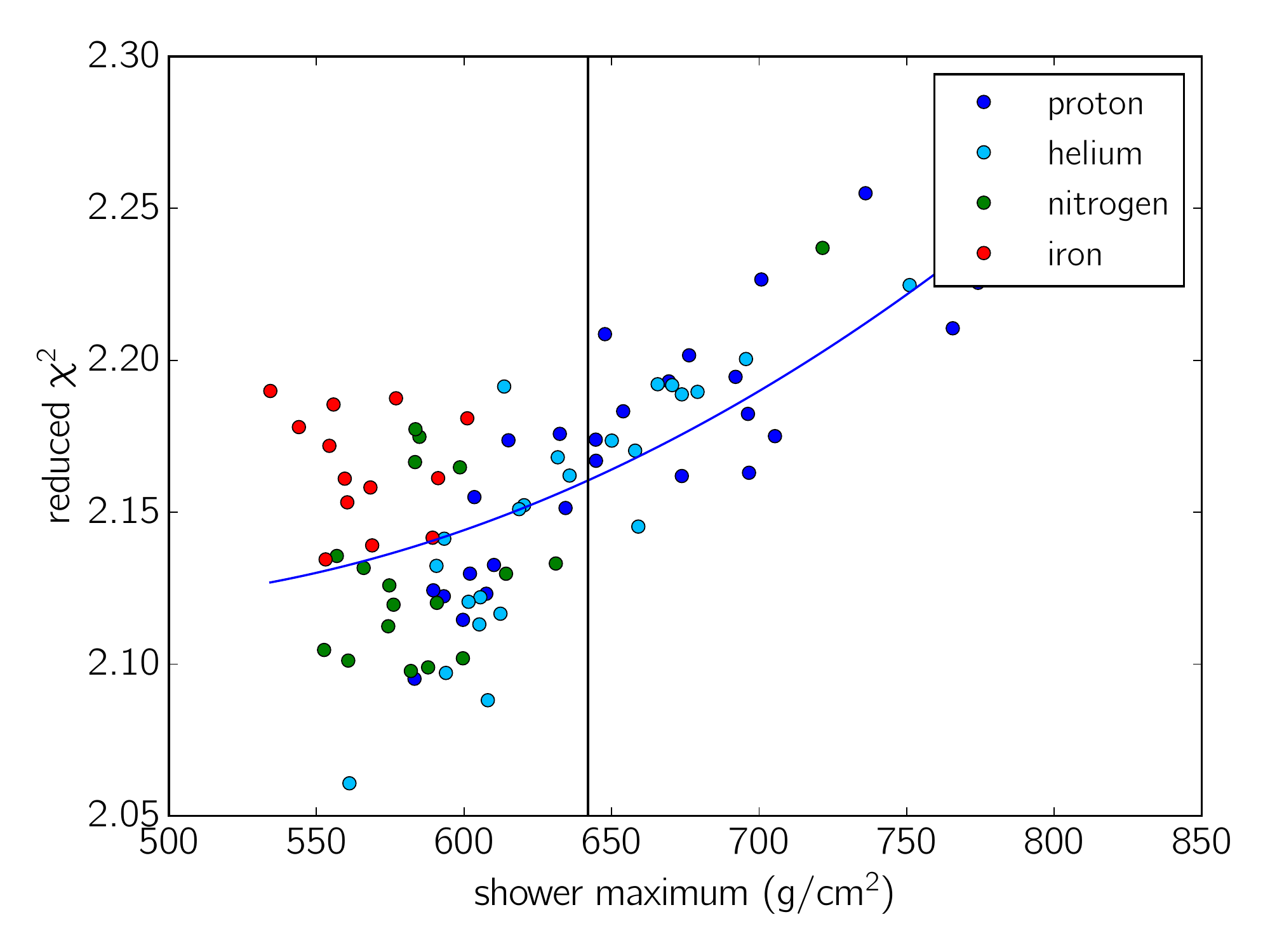}\\
\caption{Example of the $X_\mathrm{max}$ distributions over reduced $\chi^2$.
The minimum of the fitted parabola is the reconstructed $X_\mathrm{max}$, 
the vertical line indicates the $X_\mathrm{max}$ reconstructed by Tunka-133.
On the right plot one can see, that the $\chi^2$ distribution is almost flat and the minimum of parabola is outside of the covered range.
This indicates that the measurement of this event is not sensitive to the shower maximum, and the event is rejected.}
\label{fig:xmax_chi}
\end{figure}

\begin{figure}[t]
\centering
\includegraphics[width=0.49\linewidth]{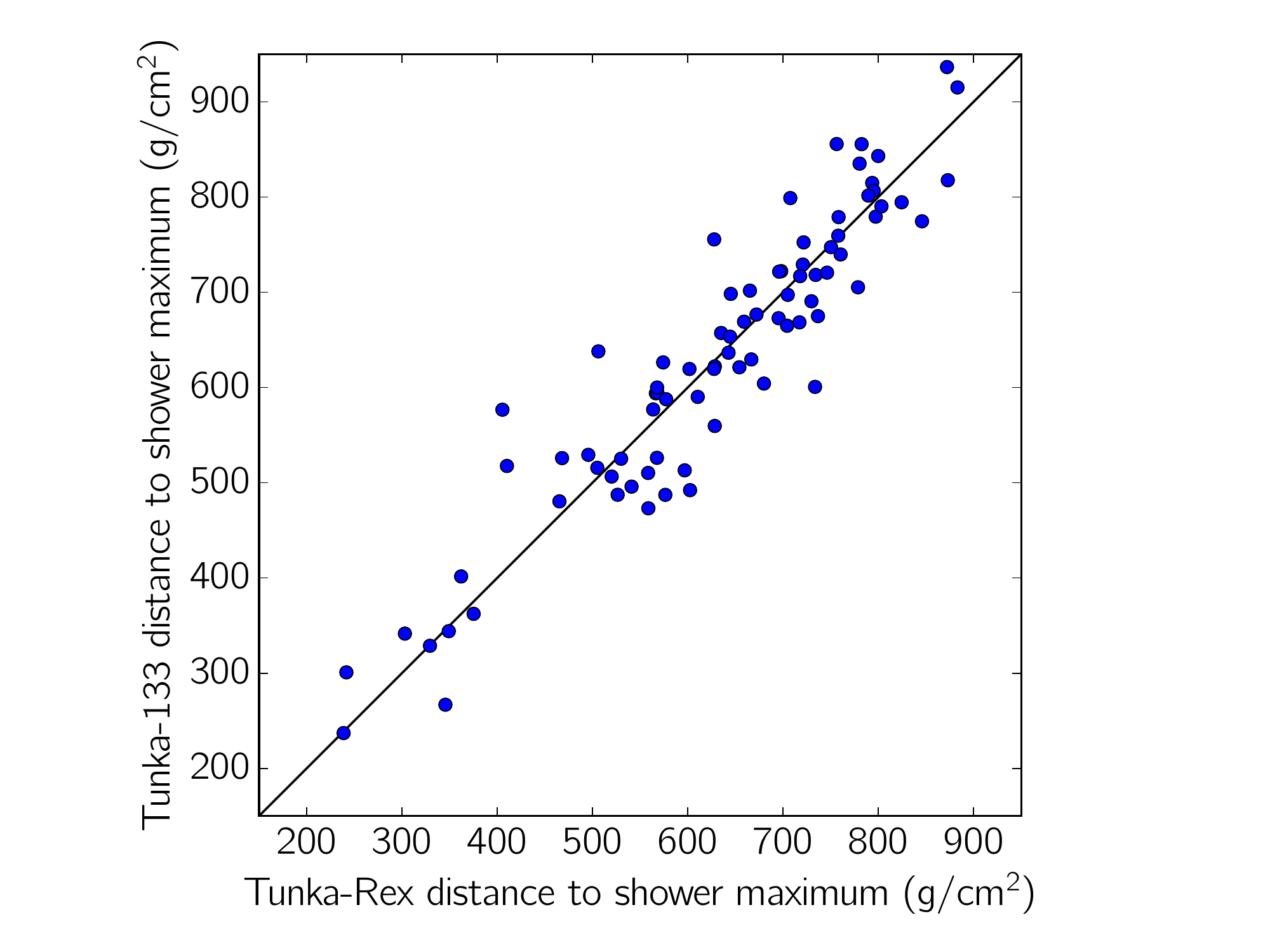}~\includegraphics[width=0.49\linewidth]{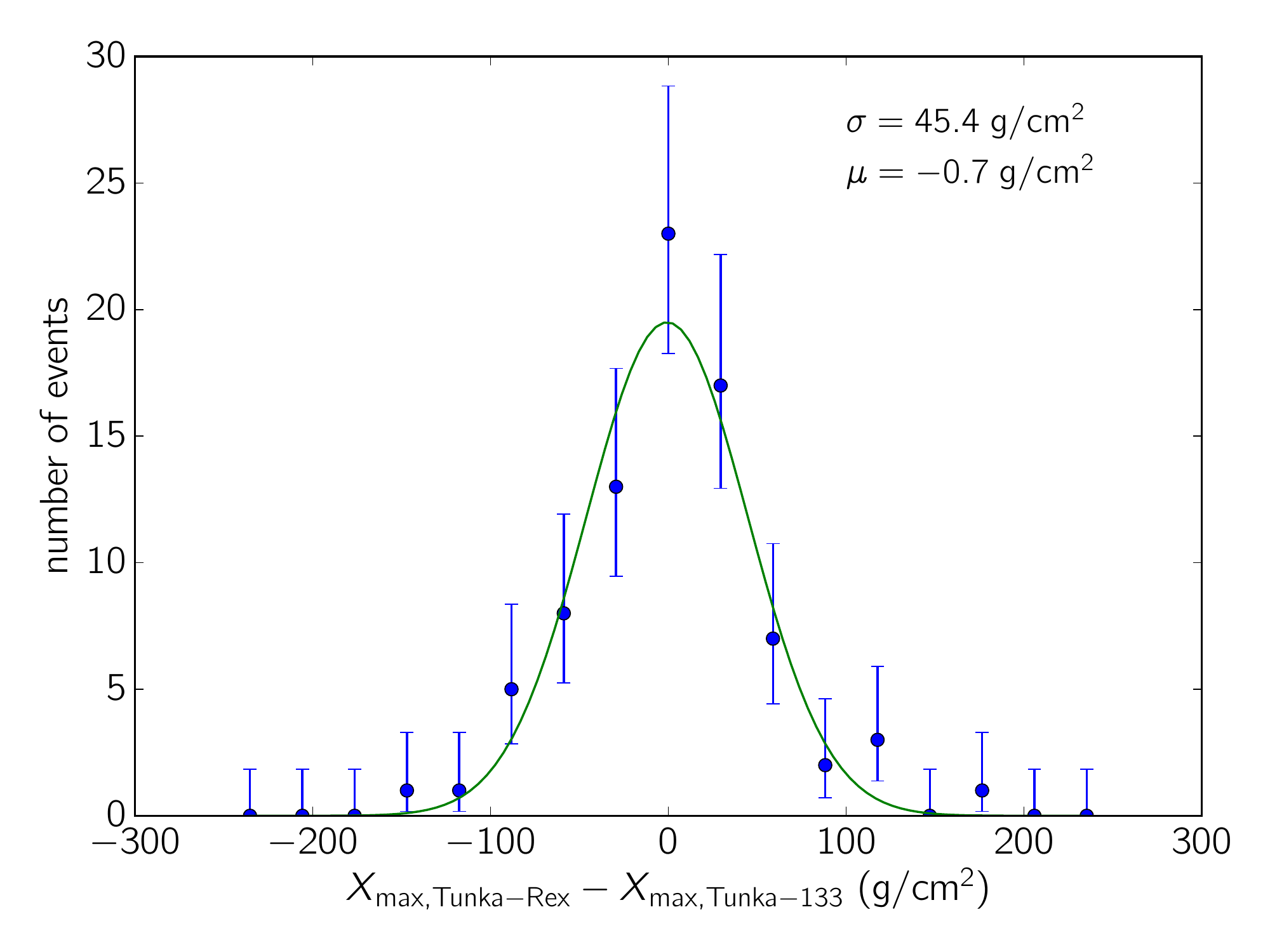}\\
\caption{Correlation and difference of the distance from the detector to $X_\mathrm{max}$ as reconstructed with Tunka-Rex radio and Tunka-133 air-Cherenkov measurements.}
\label{fig:xmax}
\end{figure}

After the minimization of $\chi^2$ for each simulation, the distribution of $\chi^2$ over the shower maximum $X_\mathrm{max}$ for is fitted each event with a parabola, 
of which the minimum is the reconstructed $X_\mathrm{max}$ (see~Fig.~\ref{fig:xmax_chi}).
We omit events, which have a reconstructed $X_\mathrm{max}$ outside of the of simulated range (see~Fig.~\ref{fig:xmax_chi}, right).
After this, we obtained twice more events with shower maximum reconstruction compared to the old Tunka-Rex analyses~\cite{Bezyazeekov:2015ica}.
Cross-check with the Tunka-133 reconstruction have shown that both experiments are in agreement in absolute scales, and the resolution of the $X_\mathrm{max}$ reconstruction with the new method is about 35~g/cm\textsuperscript{2}.
The comparison between values reconstructed by Tunka-133 and Tunka-Rex is given in Fig.~\ref{fig:xmax}.

\begin{figure}[t]
\centering
\includegraphics[width=0.49\linewidth]{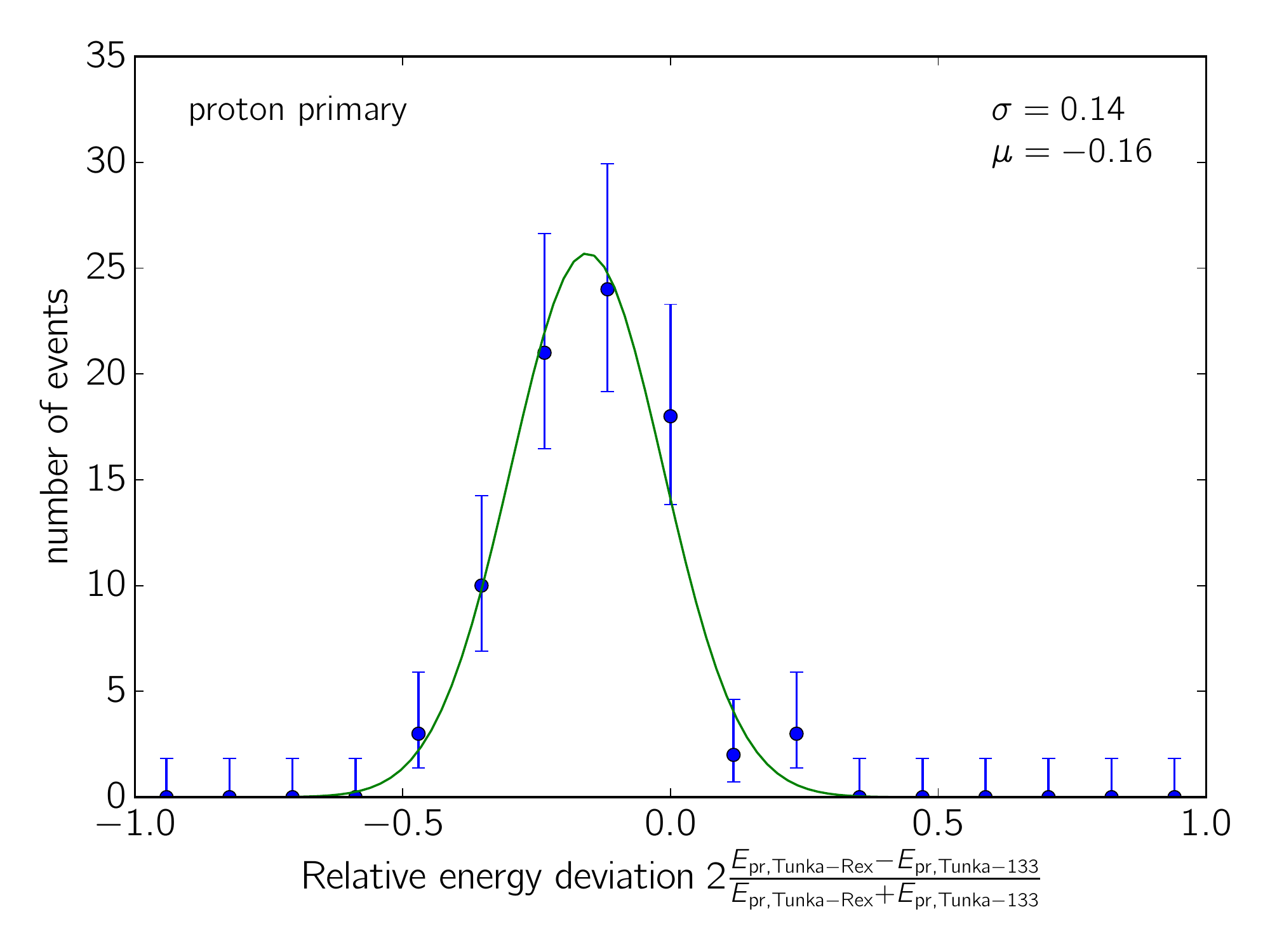}~\includegraphics[width=0.49\linewidth]{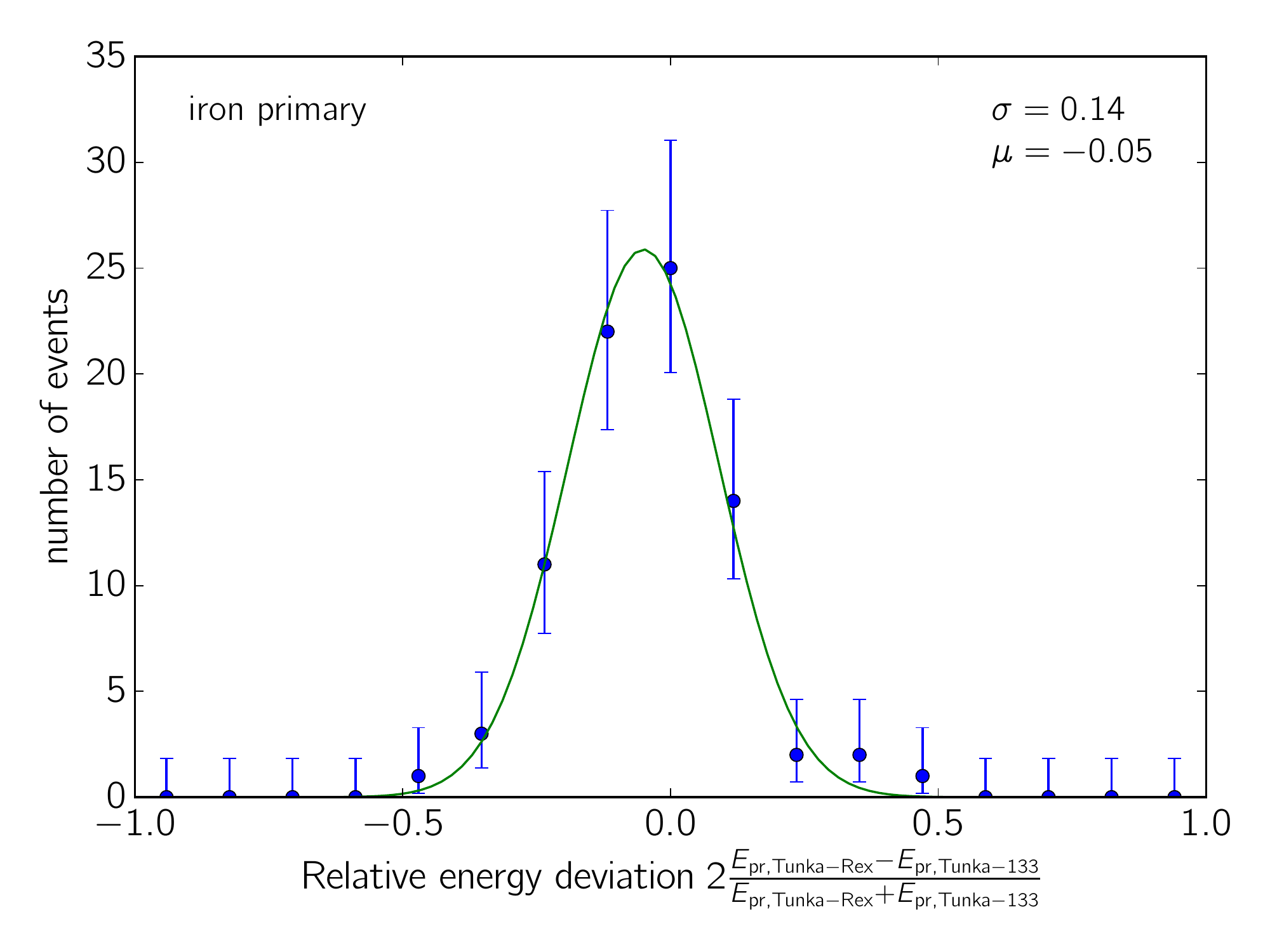}\\
\caption{Relative deviation of the shower energy reconstructed with the Tunka-133 air-Cherenkov and the Tunka-Rex radio measurements in assumption of proton (left) and iron (right) primaries.
One can see, that the iron assumption gives larger primary energy due to a smaller electromagnetic energy deposit.}
\label{fig:energy}
\end{figure}

The energy reconstruction was not the main focus of the present work, by this the preliminary reconstruction of the primary energy is performed with simple method 
of taking initial energy normalized by $A$ with minimal $\chi^2$ for events left after fitting.
Since the correction for the mass composition is in progress, this normalization is performed for each primary particle separately.
The results of the energy reconstruction in comparison with Tunka-133 is given in Fig.~\ref{fig:energy}.
Comparing to the old reconstruction we obtained systematic shift of about 10\%, which can be explained by the distortion introduced by noise, which can contribute up to 25\% of amplitude for small SNRs.
Since radio is already used for the absolute scale comparison between different cosmic-ray experiments~\cite{Apel:2016gws}, the accurate estimation of the primary energy is of special importance for the radio.

Although the development of the new method discussed in the present work is in progress, there are significant improvement of the reconstruction comparing to the old Tunka-Rex analysis.
The summary of performance of both methods\footnote{In the present analysis the phase correction mentioned in Ref.~\cite{Schroeder_TunkaRex_ICRC2017} is not implemented.} is given in Table~\ref{tab:comparison}.

\begin{table}[t]
\centering
\caption{Comparison chart of the performance between the old and new Tunka-Rex analyses.
Resolutions for $E_\mathrm{pr}$ and $X_\mathrm{max}$ are derived after substraction of the Tunka-133 values of 10\% and 28~g/cm\textsuperscript{2}, respectively.}
\label{tab:comparison}
\begin{tabular}{lll}
Property            & Old method & New method\\
\hline
Number of events passing quality cuts (of 183) & 44         & 81 \\
$E_\mathrm{pr}$ resolution & 10\% & 10\% \\
$X_\mathrm{max}$ resolution & 40~g/cm\textsuperscript{2} & 35~g/cm\textsuperscript{2}\\
\end{tabular}
\end{table}

\section{Conclusion and discussion}
We made a further step in the improvement of the shower maximum reconstruction with sparse radio arrays, 
which will result in a mass composition study in the energy range of $10^{17}$~--~$10^{18}$~eV.
With the exposure accumulated in 2012-2017~\cite{Fedorov_TunkaRex_ICRC2017}, Tunka-Rex will be able to study the $\mathrm{X}_\mathrm{max}$ distribution for these energies with statistics comparable to ones already collected by existing optical detectors.

The results presented in this proceeding are still preliminary and there are important changes, which we plan to implement soon:
\begin{itemize}
\item More accurate estimation of uncertainties and biases introduced by noise. 
For the time being, as signal uncertainty we use the RMS in the noise window, however this uncertainty can be estimated more precisely, which will improve the quality of the $\chi^2$ fit.
We will also study whether the bias of about 10\% in energy reconstruction compared to the standard reconstruction is due to noise.
\item Stronger quality cuts have to be defined. 
In the present work we do not introduce any post-reconstruction quality cuts, however with increased statistics such quality cuts can be used to select events reconstructed with higher accuracy.
\end{itemize}

Generally, there is space for the further improvements of the method.
Simulations shows, that the resolution of the detector under realistic conditions can be at least 25~g/cm\textsuperscript{2} with noise and 15~g/cm\textsuperscript{2} without noise.
It is possible, that whitening the noise, e.g. with a median filter, and more accurate tuning of the matched filtering and quality cuts could improve the reconstruction.
On the other hand, there are more complex uncertainties, e.g. by atmosphere conditions~\cite{Corstanje:2017djm}.

One of the main advantages of this method is the reconstruction the air-shower parameters exploiting the pulse shape informationa and not only one point per antenna stations (amplitude or power).
Using the shower geometry of the triggering detector, not only the energy can be reconstructed from a single antenna station~\cite{Hiller:2016opd}, but also $X_\mathrm{max}$, which lowers the threshold and increases the aperture of radio extensions to particle detector arrays.

\section*{Acknowledgements}
The construction of Tunka-Rex was funded by the German Helmholtz association and the Russian Foundation for Basic Research (grant HRJRG-303).
This work has been supported by the Helmholtz Alliance for Astroparticle Physics (HAP),
by Deutsche Forschungsgemeinschaft (DFG grant SCHR 1480/1-1),
by the Russian Federation Ministry of Education and Science (projects 14.B25.31.0010, 2017-14-595-0001, 3.9678.2017/BCh, N3.904.2017/PCh),
by the Russian Foundation for Basic Research (grants 16-02-00738, 16-32-00329, 17-02-00905),
and by grant 15-12-20022 of the Russian Science Foundation (section~4). \bibliographystyle{JHEP}
\bibliography{references}

\end{document}